\documentclass[%
reprint,
 amsmath,amssymb,
 aps,
]{revtex4-2}

\usepackage{graphicx}
\usepackage{dcolumn}
\usepackage{bm}

\begin{document}


\title{Non-Gaussian Effects of the Saha's Ionization in the Early Universe}

\author{L. L. Sales $^1$}
\email{lazarosales@alu.uern.br}
\author{F. C. Carvalho $^1$}
\email{fabiocabral@uern.br}
\author{E. P. Bento $^{1,2}$}
 \email{eliangela.bento@academico.ufpb.br}
\author{H. T. C. M. Souza $^3$}%
\email{hidalyn.souza@ufersa.edu.br
}
\affiliation{ $^1$Departamento de F\'\i sica, Universidade do Estado do Rio Grande do Norte, 59610--210, Mossor\'o -- RN, Brazil\\
$^2$ Departamento de F\'\i sica, Universidade Federal da Para\'\i ba, 58059--970, Jo\~ao Pessoa -- PB, Brazil \\
$^3$ Departamento de Ci\^encias Exatas e Naturais, Universidade Federal Rural do Semi--\'Arido, 59900--000, Pau dos Ferros -- RN, Brazil}

\date{\today}

\begin{abstract} 
Tsallis' thermostatistical has received increasing attention due to its success in describing phenomena that manifest unusual thermodynamic properties. In this context, the generalized Saha equation must follow a condition of generalized thermal equilibrium of matter and radiation. The present work aims to explore the non-Gaussian effects on Saha's ionization via Tsallis statistics. To accomplish this, we generalized the number density taking into account a non-Gaussian Fermi-Dirac distribution, and then set out the Saha equation for the cosmological recombination. As a result, we highlight two new non-Gaussian effects: $i$) two generalized chemical equilibrium conditions, one for the relativistic regime and the other for the non-relativistic one; and $ii$) the hydrogen binding $q$-energy. We demonstrated that to yields smooth shifts in the binding energy, the $a$-parameter must be very small. We also showed that binding $q$-energy exhibits symmetrical behavior around the value of the standard binding energy. Besides, we used the $q$-energy in order to access other hydrogen energy levels, and we ascertained the values of the $a$-parameter that access those levels and their relationship to temperature. Finally, we employed these results to examine the non-Gaussian effects of the deuterium bottleneck, recombination and the particle anti-particle excess. 
\end{abstract}

\maketitle


\section{Introduction} 

Developed by the Indian astrophysicist Meghnad Saha in 1920 \cite{Saha}, the Saha equation plays an important role in the study of the fraction of ionized atoms as a function of particle densities and temperature. Among several problems in physics, we highlight those whose determination of the ionization fraction provides vital information for the understanding of various puzzles, such as those related to the production of neutrinos in the solar core, among others related to astrophysics and cosmology. In this context, processes that deal with the formation of the primordial light elements and the formation of the neutral hydrogen atoms can be studied by the Saha equation. It is well known that the Saha equation applied to primordial nucleosynthesis and recombination provides important physical information as long as the chemical equilibrium is maintained \cite{peebles1993principles,dodelson2003modern,Piattella}. However, Saha's approach does not give an entire description of these epochs since the chemical equilibrium is broken throughout the evolution of the early universe. Therefore, it is natural to explore alternative approaches to this issue in order to evaluate the effects on the quantities of physical interest. Particularly is important to investigate the cosmological recombination via non-Gaussian statistics which implements generalized forms to Boltzmann-Gibbs Statistical Mechanics (BG).

In recent decades, there has been meaningful interest in the study of properties of physical systems that cannot be fully explained by BG statistics \cite{Abe}. In general, these systems have some specific features like memory effects, long-range interaction, self-organization, and fractal or multifractal evolution. In 1988, inspired by multifractal systems, C. Tsallis purposed a possible generalization of BG entropy \cite{Tsallis}. The so-called $q$-entropy was set out as
\begin{eqnarray}
	\label{qentropy}
	S_{q} = \frac{k_{B}}{q-1}\left(1-\sum_{i=1}^{\Omega}p_{i}^{q}\right)~,
\end{eqnarray}
where $k_{B}$ represents the Boltzmann constant, $p_{i}$ the normalized probability of the system in the microstate $i$, $\Omega$ the total number of settings, and $q$ the parameter that characterizes the degree of non-extensivity of the system. In the limit, $q \rightarrow 1$, the BG entropy is recovered. Mathematically speaking, $S_q$ is $i$)  concave for $q>0$ and $ii$) convex for $q<0$. Since classical entropy is concave for an equilibrium state, from the thermodynamic viewpoint, the $q$-entropy has a physical meaning only for positive values of the $q$-parameter.

This new non-Gaussian theory redefines the logarithm and exponential functions as
\begin{equation}
	\ln_q(x)=\frac{x^{1-q}-1}{1-q}
\end{equation}
and
\begin{equation}
	\exp_q(x)=e_{q}^{x}= {[1+(1-q) x]}^{1/{(1-q)}}~,
\end{equation}
respectively. When $q\rightarrow 1$ these functions return to standard form.

Many applications using a power-law like distribution have been observed in literature, for example, in plasma physics \cite{PhysRevE.61.3260}, cosmic rays \cite{BECK2004173}, astrophysics \cite{plastino1993stellar}, cosmology \cite{bernui2006temperature,Bertulani}, among others. On the other hand, few studies on the generalization of the Saha's ionization using non-Gaussian statistics have been carried out in recent years. Previous researches dealing with this can be found in Refs \cite{pessah2001statistical,soares2019non}. However, no non-Gaussian effect on binding energy has been reported so far. Furthermore, both approaches present final expressions for the Saha equation as a function of chemical potentials, which characterizes a difficulty that does not turn up in the standard case, once it is not common to find out $\mu$-values. Motivated by this issue, this paper aims to investigate the non-Gaussian effects of the Saha's ionization in the early universe. Specifically, we focus on primordial hydrogen recombination and Big Bang nucleosynthesis (BBN). In our approach, we start the non-Gaussian analysis from the equilibrium condition arising from the Boltzmann equation for annihilation. In this case, we shall generalize the particle number density via Tsallis statistics, establish new non-Gaussian chemical equilibrium conditions and deduce the generalized Saha equation.

This paper is organized as follows. In Section \ref{2section}, we present the standard Saha's approach to describe the neutral hydrogen formation process. Our proposal for the non-Gaussian Saha's ionization is introduced in Section \ref{3section}. Applications are presented in Section \ref{app}, where we explore the non-Gaussian effects of the deuterium bottleneck, recombination, and the particle anti-particle excess. Lastly, the conclusions are shown in Section \ref{conclusion}. Throughout the paper, we will adopt the following system of natural units: $c=k_{B}=\hbar=1$.   

\section{Saha's ionization} \label{2section}

The physical process known as Recombination is represented by photoionization reaction $e^{-}+p\leftrightarrow H+\gamma$ in which electrons and protons combine to form the neutral hydrogen atom in the early universe. A possible way to study this reaction is through the Saha equation to fractional ionization $x_{e} = n_{e}/ n_{b}$. In the limit to non-relativistic particles, i.e. $m_{i}\gg T$, the free electrons fraction is given by
\begin{eqnarray} \label{Saha-eq}
	\frac{x_{e}^{2}}{1-x_{e}} = \frac{1}{n_{b}\lambda_e^{3/2}}e^{-\varepsilon_{0}/T}~,
\end{eqnarray}
where $\lambda_{e}=2\pi/(m_{e}T)$ is the thermal electron wavelength and $\varepsilon_{0}=m_{e}+m_{p}-m_{H}=13.598~\textrm{eV}$ the ground state binding energy of the hydrogen atom. Further, $T$ is the temperature of the thermal bath and $m_{e}$ the electron mass. Here, $n_{b}=n_{e}+n_{H}$ is the baryons density determined from baryon-to-photon ratio  
by \cite{dodelson2003modern}
\begin{eqnarray}
	n_{b}=\eta_{b}\frac{2\zeta(3)}{\pi^2}T^3~.
\end{eqnarray}
In the expression above, $\zeta(z)$ is the Riemann's zeta function and $\eta_b=2.75 \times 10^{-8}\Omega_{b0}h^{2}$, where $\Omega_{b0}$ is the baryonic matter density parameter in the current time and $h$ the Hubble constant. The cosmological redshift is related to temperature by $T=T_0(1+z)$.

In fact, when $T\approx \varepsilon_{0}$, we obtain from Eq. ($\ref{Saha-eq}$) that $x_e \approx 1$. In other words, even when the energy of the thermal bath drops below the binding energy of the hydrogen atom, still no hydrogen is formed and the electrons remain free. This happens because the high photon-to-baryon number delays recombination, destroying any newly formed neutral atom. It is straightforward to show using a numerical calculation that the free electron fraction falls sharply at about $T\approx 0.30~\textrm{eV}$. Hence, the Saha equation works until chemical equilibrium is kept. Assuming homogeneous distributions of matter and radiation the equilibrium fractional ionization, i.e. $x_e=0.5$, provides $T_{\rm dec}\approx 0.321~\textrm{eV}$ which corresponds to a redshift $z_{\rm dec}\approx 1360$ \cite{peebles1993principles}.

\section{Non-Gaussian Saha's Ionization} \label{3section}

\subsection{Preliminary}

In classical statistics, the recombination is studied via the Saha equation from the equilibrium condition. Therefore, for generalizes the Saha equation in the Tsallis framework, we will consider the generalized version of this equilibrium state as follow:
\begin{eqnarray} \label{q-equilibrium}
	\frac{n_{e}^{q}n_{p}^{q}}{n_{H}^{q}} = \frac{n_{e}^{(0),q}n_{p}^{(0),q}}{n_{H}^{(0),q}}~.
\end{eqnarray}

In this paper, we will examine the non-Gaussian effects on the left-hand side of Eq. (\ref{q-equilibrium}). For this case, the non-Gaussian fractional ionization may be written as
\begin{eqnarray} \label{q-Saha}
	\frac{(x_{e}^{q})^2}{1-x_{e}^{q}} = \frac{n_{e}^{q}n_{p}^{q}}{n_{H}^{q}n_{b}^{q}}~,
\end{eqnarray}
where $x_{e}^{q} = n_{e}^{q}/n_{b}^{q}$ stands for the generalized degree of ionization and $n_{b}^{q}=n_{e}^{q}+n_{H}^{q}$ the baryons $q$-density. In general, $n_{i}^{q}$ can be defined in the momentum space as
\begin{eqnarray} \label{dnpeg} 
	n_{i}^{q} =\frac{g_{i}}{(2\pi)^{3}}\int d^{3}p \mathcal{N}_{i}^{q}~,
\end{eqnarray}
where $\mathcal{N}_{i}^{q}$ is the generalized occupation number and $g_i$ the degeneracy of the species. The non-Gaussian quantum statistics provide generalized forms of distribution to fermions and bosons. Since we are dealing with non-relativistic fermions, we will adopt the following fermionic distribution \cite{buyukkilic:1993bd}
\begin{eqnarray} \label{gon}
	\mathcal{N}_{i}^{q} = \frac{1}{e_{2-q}^{\beta (E_{i}-\mu_{i})}+1}~,
\end{eqnarray}
with $E_i$ and $\mu_i$ being the energy and chemical potential of the species $i$, respectively. Besides, $e_{2-q}^{x} = [ 1+(q-1)x]^{1/(q-1)}$ and $\beta=1/T$. 

We are interested in systems with temperatures less than $E_{i}-\mu_{i}$, because in this limit we can neglect the quantum effects since  $e_{2-q}^{\beta (E_{i}-\mu_{i})} \gg 1$. Thus, Eq. (\ref{gon}) becomes 
\begin{eqnarray} 
	\mathcal{N}_{i}^{q} = \left[1+(1-q)\beta (\mu_{i}-E_{i})\right] ^{\frac{1}{1-q}} = e_{q}^{\beta (\mu_{i}-E_{i})}~.
\end{eqnarray} 
Using a $q$-algebra space like given in \cite{borges2004possible}, we were able to rewrite Eq. (\ref{dnpeg}) in the non-relativistic limit as
\begin{eqnarray} \label{dnpeg1}
	n_{i}^{q} = \frac{4\pi g_{i}}{(2\pi)^{3}}e_{q}^{\beta(\mu_{i}-m_{i})}\int_{0}^{a} p^{2}\left[1+(q-1)\gamma_{i} p^2\right]^{\frac{1}{1-q}}dp~,
	\nonumber\\
\end{eqnarray}
where 
\begin{eqnarray} \label{a}
	a = \left\{
	\begin{array}{rl}
		\infty, & q>1 \\
		\left[(1-q)\gamma_{i}\right] ^{-1/2}, & q<1~,
	\end{array}
	\right.
\end{eqnarray}
with factor $\gamma_{i}$ given by
\begin{eqnarray} 
	\gamma_{i} = \frac{\beta}{2m_{i}[1+(1-q)\beta(\mu_{i}-m_{i})]}~.
\end{eqnarray}
Evaluating the integral in Eq. (\ref{dnpeg1}), we can write the particle number $q$-density for $q>1$ and $q<1$ as
\begin{eqnarray} \label{q-density}
	n_{i}^{q} = \frac{g_{i}B_{q}}{\lambda_i^{3/2}}\left[ e_{q}^{\beta(\mu_{i}-m_{i})}\right]^{\frac{5-3q}{2}}~,
\end{eqnarray}
where
\begin{eqnarray} 
	B_{q} = \left\{
	\begin{array}{rcl}
		\displaystyle{\frac{1}{(q-1)^{3/2}}\frac{\Gamma\left(\frac{5-3q}{2(q-1)}\right)}{\Gamma\left(\frac{1}{q-1}\right)}},& \mbox{if} & 1<q<5/3 \\ \\
		\displaystyle{\frac{1}{(1-q)^{3/2}}\frac{\Gamma\left(\frac{2-q}{1-q}\right)}{\Gamma\left(\frac{7-5q}{2(1-q)}\right)}}, & \mbox{if} & q<1~.
	\end{array}
	\right.
\end{eqnarray}
In the limit $q\rightarrow 1$, we have that $B_q\rightarrow 1$, $e_{q}^{x}\rightarrow e^{x}$ and then the standard particle number density is recovered. Having determined the non-Gaussian shape of the particle number density, we can proceed to the next step which is the deduction of the generalized Saha equation. But before that, let us determine the non-Gaussian chemical equilibrium conditions.

\subsection{Non-Gaussian chemical equilibrium conditions}

Here, we aim to check if there are non-Gaussian effects on the chemical equilibrium condition. To accomplish this, we use Eq. (\ref{q-density}) to express 
\begin{eqnarray} \label{nin0q}
	\frac{n_{i}^{q}}{n_{i}^{(0),q}} = \left( e_{q}^{\theta_{i}^{q}} \right)^{\frac{5-3q}{2}} \;,
\end{eqnarray}
where
\begin{eqnarray}
	\theta_{i}^{q} = \frac{\beta\mu_{i}}{1-(1-q)\beta m_{i}}\;,
\end{eqnarray}
and replace Eq. (\ref{nin0q}) in Eq. (\ref{q-equilibrium}) to obtain 
\begin{eqnarray} \label{theta_eq}
	\theta_{H}^{q} = \theta_{e}^{q} + \theta_{p}^{q} + (1-q)\theta_{e}^{q}\theta_{p}^{q}\;.
\end{eqnarray}

Considering $m_{H}\approx m_{p}$ in Eq. (\ref{theta_eq}), the non-Gaussian chemical equilibrium for non-relativistic particles shall be written as
\begin{eqnarray} \label{ceqg}
	\mu_{H} = \mu_{p} + \sigma_{q}\mu_{e}[1 + (1-q)\beta\mu_{p}]\;,
\end{eqnarray}
where 
\begin{eqnarray} 
	\sigma_{q} = \frac{1-(1-q)\beta m_{p}}{1-(1-q)\beta m_{e}}\;.
\end{eqnarray}
Let us now ascertain a condition such that chemical equilibrium for relativistic particles is reached. In this limit,  $T\gg m_{i}$, which leads to $\sigma_{q}\approx 1$ and for that reason Eq. (\ref{ceqg}) becomes  
\begin{eqnarray} \label{ceqgr}
	\mu_{H} = \mu_{p} + \mu_{e} + (1-q)\beta\mu_{e}\mu_{p}\;.
\end{eqnarray}
Hence, the suitable shapes for the chemical equilibrium in the Tsallis framework are given by Eqs. (\ref{ceqg}) and (\ref{ceqgr}). It is worth noting that the chemical equilibrium condition in classical statistics works for both relativistic and non-relativistic limits, whereas in the Tsallis framework there is a condition for each regime, as we have just demonstrated. The chemical $q$-equilibrium for non-relativistic particles presents a mass dependence, it does not turn out in the standard case. According to Eq. (\ref{ceqg}), even that $T\gg \mu_{e}\mu_{p}$, we still have $\mu_{H} = \mu_{p} + \sigma_{q}\mu_{e}$. In this case, for the usual chemical equilibrium to be achieved, we must set $q=1$. In contrast, for relativistic particles, the standard chemical equilibrium is recovered in two situations: $i)$ when $q=1$ or $ii)$ whenever $T\gg \mu_{e}\mu_{p}$, namely $\mu_{e}+\mu_{p}=\mu_{H}$.

\subsection{Non-Gaussian Saha equation} 

In order to investigate the ionization process in the non-Gaussian context, we analyze the reaction in Eq. (\ref{q-equilibrium})  
by capturing a free-electron directly to the ground state and we model the process like a classic gas with  $\mathcal{N}_{i}^{q}\ll 1$ \cite{peebles1993principles}. Taking $n_{i}^{q}$ for electrons ($e^{-}$), protons ($p$) and neutral hydrogen atom ($H$), and using $g_{e}=g_{p}=g_{H}/2=2$ and $m_{H}\approx m_{p}$, Eq. (\ref{q-Saha}) may be write as 
\begin{eqnarray} \label{q-Saha1}
	\frac{(x_{e}^{q})^{2}}{1-x_{e}^{q}}=\frac{1}{n_{b}^{q}\lambda_e^{3/2}}B_{q}\left[ e_{q}^{-\beta\varepsilon_{q}}e_{q}^{\zeta_q}\right]^{\frac{5-3q}{2}}~,
\end{eqnarray}
where
\begin{eqnarray} \label{xi2}
	\zeta_q=\frac{\theta_{e}^{q} + \theta_{p}^{q} + (1-q)\theta_{e}^{q}\theta_{p}^{q}-\theta_{H}^{q}}{1+(1-q)\theta_{H}^{q}}\;.
\end{eqnarray}
Applying the condition imposed by Eq. (\ref{theta_eq}), we get $\zeta_q=0$ and hence the second $q$-exponential factor on the right-hand side of (\ref{q-Saha1}) is equal to $1$. Thus, the non-Gaussian Saha equation has the form
\begin{eqnarray} \label{q-Saha3}
	\frac{(x_{e}^{q})^2}{1-x_{e}^{q}} = \frac{25}{19n_{b}^{q}\lambda_e^{3/2}}B_{q}\left( e_{q}^{-\beta\varepsilon_{q}}\right)^{\frac{5-3q}{2}}~.
\end{eqnarray}
This result takes into account about $24\%$ of the helium fraction, i.e., $0.76n_{b}^{q}=n_{e}^{q}+n_{H}^{q}$. One of the advantages of our approach is that Eq. (\ref{q-Saha3}) does not depend on chemical potentials, which makes it stand out from other ones found in the literature. 

\begin{figure}[ht!] 
	\centering
	\includegraphics[scale=0.3]{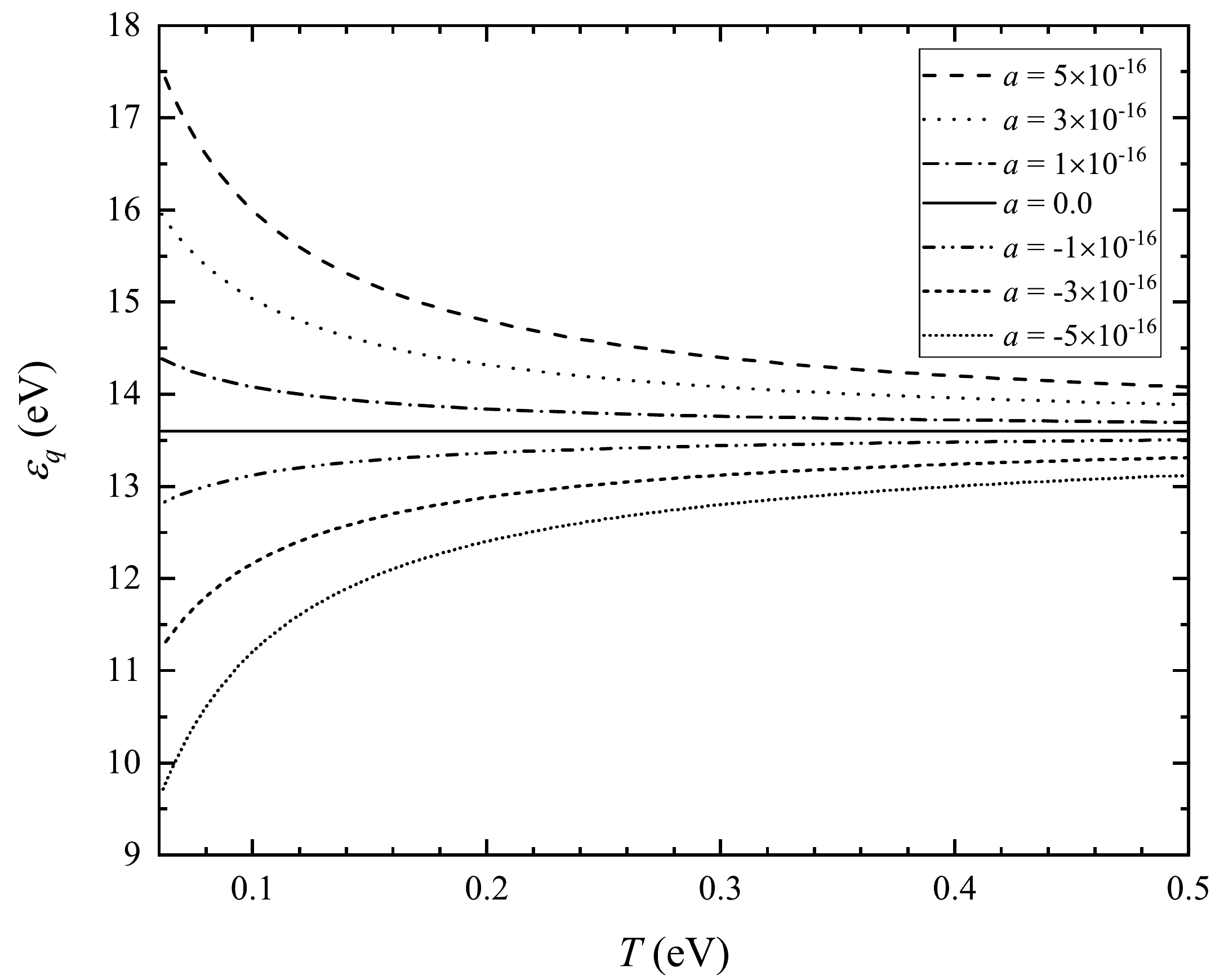}
	\caption{The behavior of non-Gaussian binding energy as a function of temperature for some values of the $a$-parameter. This result takes into account that $m_{e}=5.110\times 10^{5}~\rm{eV}$, $m_{p}=9.383\times 10^{8}~\rm{eV}$, $m_{H}=9.389 \times 10^{8}~\rm{eV}$ and $\varepsilon_{0}=13.598~\rm{eV}$ \cite{Zyla:2020zbs}.}
	\label{fig1}
\end{figure}

Unlike the conventional approach, the binding energy in the Tsallis framework has a dependency on the $q$-parameter and the temperature. We define the binding $q$-energy as follows:  
\begin{eqnarray} \label{gbe}
	\varepsilon_{q} =\frac{\varepsilon_{0}+(q-1)\beta m_{e}m_{p}}{1+(q-1)\beta m_{H}}~. 
\end{eqnarray} 
Note that the usual binding energy is also recovered in two situations: $i)$ when $q\rightarrow 1$ or $ii)$ when $T\gg m_{e}m_{p}$, i.e., $\varepsilon_{q} \rightarrow \varepsilon_{0}$. Figure 1 shows the behavior of binding $q$-energy as a function of temperature according to some values of the $a$-parameter. For reasons of simplicity, we set out $a=q-1$. For $a\neq 0$, an asymptotic behavior appears as a function of temperature. The curves also present symmetric behavior for the values of the $a$-parameter; for $a>0$, we have $\varepsilon_{q}>\varepsilon_{0}$ and for $a<0$, $\varepsilon_{q}<\varepsilon_{0}$. At high temperatures, mainly for tiny changes in $a$, the measure of  $\varepsilon_{q}$ approaches the binding energy in the ground state. In other words, at very high temperatures, $T\rightarrow \infty$, the non-Gaussian effect on binding energy is suppressed. For temperatures near zero, $\varepsilon_{q}$ becomes too large for $a>0$ and too small when $a<0$. Besides, figure \ref{fig1} also displays that the curves are more sensitive to the $a$-parameter at low temperatures than at high ones. In summary, we showed that to yield smooth shifts in the binding energy, the $a$-parameter must be very small, i.e., $a\sim 10^{-16}$. This means that the non-Gaussian effect on binding energy is slight when $a$ is small and rises quickly as $a$ grows up mostly at low temperatures. 

Once the non-Gaussian effect on binding energy is a ground state shift, we can use that to figure out which values of the $a$-parameter allow accessing other energy levels of the hydrogen atom. Taking $\varepsilon_{q,n} = \varepsilon_{0}/n^2$, Eq. (\ref{gbe}) leads to 
\begin{eqnarray} \label{eqpt}
	a = \frac{\varepsilon_{0}\left(1 - \frac{1}{n^2}\right)}{\beta\left(\frac{\varepsilon_{0}}{n^2}m_H-m_{e}m_p\right)}~.
\end{eqnarray}
Note that $\varepsilon_{q,n}$ selects the desired energy level of the hydrogen atom as long as $n\geq 1$. As $m_{e}m_p\gg \varepsilon_{0}m_H/n^2$, then for accessing such energy levels we must have $a<0$. In the ground state ($n=1$), we have that $a=0$, as seen in Eq.(\ref{gbe}) for $q=1$. For excited states ($n\geq 2$), the $a$-parameter depends on both the temperature and the chosen energy level. Figure 2 shows the behavior of $a$, from Eq. (\ref{eqpt}), as a function of temperature for energy levels of the hydrogen atom. In this plane is possible to see that the $a$-parameter has descending linear behavior, changing subtly at low temperatures and highlighting their dependence over $a$ values as the temperature rise in each corresponding $n>2$ energy level. As emphasized before, the $a$-parameter remains constant with temperature in the ground state.     
\begin{figure}[ht!] 
	\centering
	\includegraphics[scale=0.3]{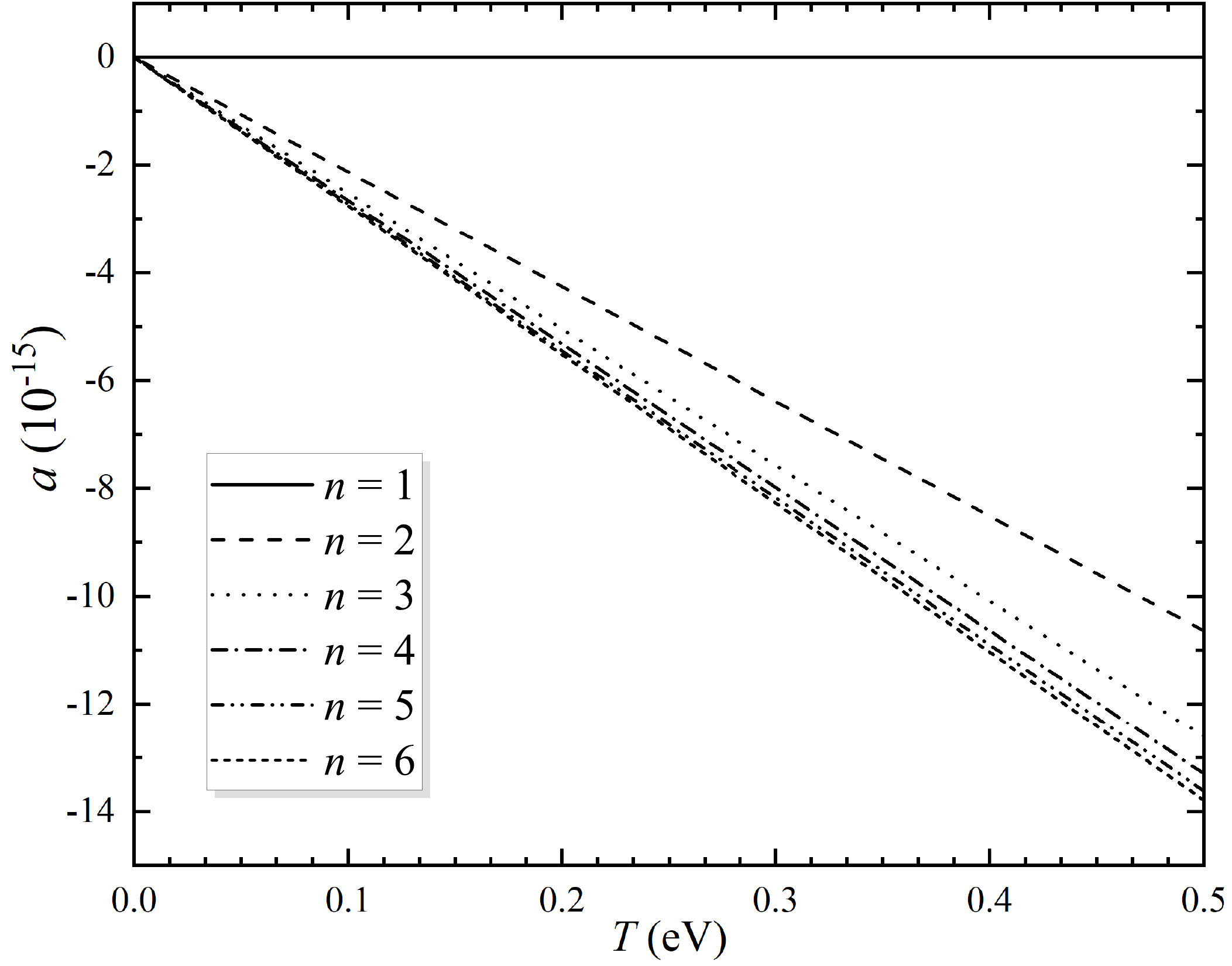}
	\caption{The evolution of the $a$-parameter with temperature for some energy levels of the hydrogen atom according to Eq. (\ref{eqpt}).}
	\label{fig2}
\end{figure}

\subsection{Determination of the baryons $q$-density} 

To obtain a full version of the non-Gaussian Saha equation, we need to specify the baryons $q$-density ($n_{b}^{q}$). To perform this, we will use the definition of baryon-to-photon ratio, as below:
\begin{eqnarray} \label{eta}
	\eta_{b}^{q} = \frac{n_{b}^{q}}{n_\gamma^{q}}~.
\end{eqnarray}
It is well known that the baryon-to-photon ratio is constant \cite{dodelson2003modern}. From Eq. (\ref{eta}) it is straightforward show that $\eta_{b}^{q}=\rho_{c0}\Omega_{b0}^{q}/(m_{p}n_{\gamma 0}^{q})$ is constant, since $\Omega_{b}^{q}=\Omega_{b0}^{q}(1+z)^3$ and $n_{\gamma}^{q}=n_{\gamma 0}^{q}(1+z)^3$. Wherefore, we can reasonably assume that $\eta_{b}^{q}\approx \eta_b=2.75 \times 10^{-8}\Omega_{b0}h^{2}$ so that Eq. (\ref{eta}) can be written as
\begin{eqnarray}
	n_{b}^{q} = 2.75 \times 10^{-8}\Omega_{b0}h^{2}n_{\gamma}^{q}~.
\end{eqnarray}

However, for photons $\mu_{\gamma}=0$, $g_\gamma=2$, $E=p$, and thus, the photons $q$-density according to Eq. (\ref{dnpeg}) becomes
\begin{eqnarray} \label{intq}
	n_{\gamma}^{q} = \frac{8\pi}{(2\pi\beta)^{3}}\int_{0}^{\infty}\frac{x^{2}dx}{e_{2-q}^{x}-1}\;,
\end{eqnarray}
where we defined $x=\beta p$. More details about the exact solution of this expression can be consulted in \ref{appendixA}. Hence, Eq. (\ref{intq}) is given by
\begin{eqnarray} \label{q-number-pho}
	n_{\gamma}^{q}  = \frac{C_{q}}{\pi^{2}}T^{3}
\end{eqnarray}
with $C_q$ being
\begin{eqnarray}
	C_q = \frac{2\psi^{(0)}(3-2q)-\psi^{(0)}(4-3q)-\psi^{(0)}(2-q)}{(q-1)^2},
	\nonumber\\
\end{eqnarray}
where $\psi^{(n)}(z)$ is a special function so-called polygamma function or most commonly referred to by $n$-th derivative of the digamma function. Since this function diverge when $z=0$, the $q$-parameter is constrained as follow: $q\neq 4/3, q\neq 3/2 \;{\rm and}\; q\neq 2$. Most specifically, we have three likely schemes: $0<q<4/3$, $4/3<q<3/2$, and $3/2<q<2$. But the second scenario yields a negative result for photon density, and this is not physically acceptable. Moreover, we have that $C_q>0$ if $q<0$, but is well known that $q$-entropy is concave only for $q>0$, thus this picture is not suitable as well. In conclusion, the best scenario is one in which $C_q$ fluctuates around $2\zeta(3)$, i.e., $0<q<4/3$. 

In this manner, the baryons $q$-density takes the form
\begin{eqnarray}
	n_{b}^{q} = 2.75 \times 10^{-8}\Omega_{b0}h^{2}\frac{C_{q}}{\pi^{2}}T^{3}\;.
\end{eqnarray}
In the limit $q\rightarrow 1$, the factor $C_q \rightarrow 2\zeta(3)$ and then the usual baryons density is recovered. In a compact form, Eq. (\ref{q-Saha3}) may be conveniently written as
\begin{eqnarray} \label{q-saha-full}
	\frac{(x_{e}^{q})^2}{1-x_{e}^{q}} = \left(\frac{1.1\times 10^{16}}{\Omega_{b0}h^{2}}\right)D_{q}\left(\frac{T}{\rm eV}\right)^{-3/2}\left( e_{q}^{-\beta\varepsilon_{q}}\right)^{\frac{5-3q}{2}}\;,
	\nonumber\\	
\end{eqnarray}
where $D_{q} = B_{q}/C_{q}$. This is the non-Gaussian Saha equation in the reduced form. The equation above is constrained according to the interval $0<q<4/3$. The usual Saha equation is easily recovered when we take the limit $q\rightarrow 1$. For calculation purposes, we will adopt $\Omega_{b0}h^{2}=0.0225$ \cite{Planck}. 

\section{Applications} \label{app}

\subsection{The deuterium bottleneck}

The BBN was accountable for the formation of the light elements in the primordial universe. It occurred at a temperature of about $0.1~{\rm MeV}$, which corresponds to a redshift $z\approx 10^{9}$. In an environment whose temperature of the thermal bath is lower than $2.22~{\rm MeV}$ (deuterium binding energy), the photon-to-baryon ratio is still high and so the newly formed deuterium nuclei are destroyed. This is known as the deuterium bottleneck \cite{dodelson2003modern,Piattella}. 

To evaluate this bottleneck situation, we must consider the deuterium production reaction, namely $p + n \leftrightarrow D + \gamma$. The Saha equation for this physical process is given by
\begin{eqnarray} \label{q-Saha-Deuterium}
	\frac{n_D}{n_{p}n_n} = \frac{3}{4}\left(\frac{2\pi m_D\beta}{m_{p}m_{n}}\right)^{3/2}e^{\beta B_D}~, 
\end{eqnarray}
where $B_D=2.22~{\rm MeV}$ is the deuterium binding energy. For estimation purposes, we will adopt $n_b\approx 6/5n_p\approx 6n_n$ and $m_D/(m_{p}m_n)\approx 2/m_p$. Thereby, we can write the deuterium-to-baryon ratio as follow:
\begin{eqnarray} \label{d-to-b}
	\frac{n_D}{n_b} \approx \frac{5\zeta(3)}{3\sqrt{\pi}}\eta_{b}\left(\frac{T}{m_{p}}\right)^{3/2}e^{B_D/T}~. 
\end{eqnarray}  
If $T\gg B_D$ the relative deuterium abundance is entirely negligible once it is exponentially suppressed. Though, even when $T\sim B_D$ the relative abundance is still very small, because of the prefactor $\eta_b$. In other words, the prefactor $\eta_b$ inhibits nuclei production until the temperature drops well below the deuterium binding energy. 
 
From Eq. (\ref{d-to-b}) we can estimate the BBN start temperature, i.e., one at which the bottleneck is overcome. In this case, $T_{\rm BBN}$ is defined as one for which $n_D\approx n_b$ and, therefore, numerically solving Eq. (\ref{d-to-b}), we obtain $T_{\rm BBN}\approx 0.062~{\rm MeV}$. This suggests that deuterium production takes place from this temperature.

In order to explore the non-Gaussian effects of the deuterium bottleneck, we will take into account our results for both the number $q$-density and the non-Gaussian chemical equilibrium condition. We are interested in knowing the influence of the $q$-parameter on the temperature $T_{\rm BBN}$ through the generalization of Eq. (\ref{d-to-b}). Henceforward, we will carry out the same procedure adopted to obtain $T_{\rm BBN}$ in the standard case. So, the non-Gaussian version of Eq. (\ref{q-Saha-Deuterium}) can be set out as 
\begin{eqnarray}
	\frac{n_{D}^{q}}{n_{p}^{q}n_{n}^{q}} = \frac{3}{4}\left(\frac{2\pi m_D\beta}{m_{p}m_{n}}\right)^{3/2}\frac{1}{B_{q}}\left( e_{2-q}^{\beta B_{D}^{q}}\right)^{\frac{5-3q}{2}}~, 
\end{eqnarray} 
where deuterium binding $q$-energy was defined as
\begin{eqnarray} \label{Dgbe}
	B_{D}^{q} =\frac{B_{D}+(q-1)\beta m_{p}m_{n}}{1+(q-1)\beta m_{D}}~. 
\end{eqnarray}
Supposing that $n_{b}^{q}\approx 6/5n_{p}^{q}\approx 6n_{n}^{q}$, the non-Gaussian deuterium-to-baryon ratio can be written as
\begin{eqnarray}
	\frac{n_{D}^{q}}{n_{b}^{q}} \approx \frac{5}{6\sqrt{\pi}}\frac{\eta_{b}}{D_{q}}\left(\frac{T}{m_{p}}\right)^{3/2}\left( e_{2-q}^{B_{D}^{q}/T}\right)^{\frac{5-3q}{2}}~. 
\end{eqnarray}

Assuming that $n_{D}^{q}\approx n_{b}^{q}$, we can numerically estimate the $q$-parameter by solving the following equation for $T_{\rm BBN}\approx 0.062~{\rm MeV}$:
\begin{eqnarray}
	-\frac{B_{D}^{q}}{T_{\rm BBN}} &\approx&\ln_q\left(E_q\right)^{\frac{2}{5-3q}}+\ln_q\left(\frac{T_{\rm BBN}}{m_{p}}\right)^{\frac{3}{5-3q}}
	\nonumber \\
	&+&
	(1-q)\ln_q\left(E_q\right)^{\frac{2}{5-3q}}
	\ln_q\left(\frac{T_{\rm BBN}}{m_{p}}\right)^{\frac{3}{5-3q}}~, 
\end{eqnarray}
where $E_q=5\eta_{b}/(6\sqrt{\pi}D_{q})$. A meaningful outcome is that there is a value $q\neq 1$ such that $T_{\rm BBN}\approx 0.062~{\rm MeV}$. Wherefore, the deuterium bottleneck is overcome when the $q$ parameter is about $q\approx 1.152$. This result is in agreement with the mathematical constraint $0<q<4/3$ contained in the factor $E_q$. It is noteworthy that the non-Gaussian Saha equation is hybrid in the sense that it encompasses both equilibrium and non- equilibrium regimes. According to Eq. (\ref{theta_eq}), therefore, the value $q \approx 1.152$ seems to indicate that the factor $q$ is in the chemical equilibrium range in the non-Gaussian context. This is the possible physical reason for which the $q$-parameter differs from $1$.  
\begin{table}[h!] 
	\begin{center}
		\begin{tabular}{ccc}
			\hline\hline 
			$q$ & $T_{\rm BBN}~(\rm MeV)$ & $z_{\rm BBN}$ \\ \hline
			$1.103$ & $1.000$ & $4.2\times10^{9}$\\
			$1.151$ & $0.070$ & $3.0\times10^{8}$\\
			$1.152$ & $0.062$ & $2.7\times10^{8}$ \\
			$1.155$ & $0.050$ & $2.1\times10^{8}$ \\
			$1.195$ & $0.001$ & $4.2\times10^{6}$ \\
			\hline \hline
		\end{tabular}
		\caption{Estimation for $T_{\rm BBN}$ from non-Gaussian perspective. Since we estimate the factor $q$ for $T_{\rm BBN}$, in this table we estimate $T_{\rm BBN}$ from fluctuations around $q\approx 1.152$.} 
		\label{table2}
	\end{center}
\end{table}
Table \ref{table2} shows the estimate for both the temperature $T_{\rm BBN}$ and the redshift $z_{\rm BBN}$ for some values of the $q$-parameter. According to the values in Table \ref{table2}, we can conclude that $T_{\rm BBN}\propto 1/q$. This shows how the $q$-parameter affects the temperature $T_{\rm BBN}$. Then, from our estimates, the $q$-parameter plays a role in controlling the temperature of the thermal bath, since $n_\gamma \propto T^{3}\propto 1/q^{3}$. 

\subsection{Recombination}

From now on we will examine the non-Gaussian effects on primordial recombination. Using Eq. (\ref{q-saha-full}) we can estimate the equilibrium fractional ionization temperature for some values of the $q$-parameter. Table \ref{table1} presents the numerical outcomes for both temperature and redshift. 
\begin{table}[h!] 
	\begin{center}
		\begin{tabular}{ccc}
			\hline\hline 
			$q$ & $T_{\rm dec}~(\rm eV)$ & $z_{\rm dec}$ \\ \hline
			$1.094$ & $1835$ & $\sim10^{7}$\\
			$1.196$ & $1.800$ & $7664$\\
			$1.209$ & $0.321$ & $1360$ \\
			$1.210$ & $0.280$ & $1191$ \\
			$1.211$ & $0.240$ & $1021$ \\
			\hline \hline
		\end{tabular}
		\caption{Results for the equilibrium fractional ionization temperature according to some values of the $q$-parameter.}
		\label{table1}
	\end{center}
\end{table}
Similarly to the BBN case, recombination also enables a value $q\neq 1$ such that the equilibrium fractional ionization temperature is about $T_{\rm dec}\approx 0.321~{\rm eV}$. In this case, $q$-parameter is estimated to be $q\approx 1.209$. Employing the arguments aforementioned, this value also appears to be in the chemical equilibrium range in the Tsallis framework. Besides, we can note that $T_{\rm dec}$ is inversely proportional to the $q$-parameter, a behavior similar to that of temperature $T_{\rm BBN}$. One aspect to be noted is that, e.g. for $q\approx 1.1$, the equilibrium fractional ionization temperature rises abruptly compared to $T_{\rm dec}\approx 0.321~{\rm eV}$.

It is well known that the electrons recombine principally through the first excited state ($n=2$) and cascade down to the ground state by the Lyman-$\alpha$ or the {2s}–{1s} two-photon transition \cite{Peebles}. Taking this into account, we can use Eq. (\ref{eqpt}) with $n=2$ to afford bounds for the $a$-parameter in the recombination epoch. For instance, using $T_{\rm beg}\approx 0.344~{\rm eV}$ for the beginning of recombination, the $a$-parameter is estimated to be $a_{\rm beg}\approx -7.330\times 10^{-15}$. Whereas $T_{\rm end}\approx 0.210~{\rm eV}$ for the end of recombination, we have $a_{\rm end}\approx -4.474\times 10^{-15}$. On the other hand, for $T_{\rm dec}\approx 0.321~{\rm eV}$, we obtain $a_{\rm dec}\approx -6.830\times 10^{-15}$. Also, we can ascertain the value of the $a$-parameter according to Eq. (\ref{eqpt}) when $x_{e}\approx 1$. In such a situation, no hydrogen is formed, which is equivalent to $n\rightarrow \infty$. In other words, considering $\varepsilon_{0}\approx T$, we get $a=-\varepsilon_{0}^{2}/(m_{e}m_{p})\approx -3.857\times 10^{-13}$. This could be thought of as the upper limit for the $a$-parameter, with the inferior limit being $a_{\rm end}$. These results assume that Eq. (\ref{eqpt}) can in fact be set out since it is just an assumption in order to access other energy levels of the hydrogen atom. However, our goal here is to explore the non-Gaussian effects of the generalized Saha equation for recombination. 

\subsection{Particle anti-particle excess}

To examine the particle anti-particle excess in the non-Gaussian context, let us consider the following reaction
\begin{eqnarray}\label{reaction2}
	r^{+} + r^{-} \leftrightarrow \gamma + \gamma~,
\end{eqnarray}
where $r^{+}$ e $r^{-}$ represents a generic particle and its anti-particle, respectively. If this reaction remains in equilibrium, we should have $\mu_{r^{+}}=-\mu_{r^{-}}$. We aim to analyze the reaction in the non-relativistic and relativistic limits. For the non-relativistic limit ($m_{i}\gg T$), we can use our result for the number $q$-density. Thus, according to Eq. (\ref{q-density}), we have
\begin{eqnarray} 
	n_{r^{\pm}}^{q} = \frac{g_{r^{\pm}}B_{q}}{\lambda^{3/2}}\left[ e_{q}^{\beta(\mu_{r^{\pm}}-m)}\right]^{\frac{5-3q}{2}}~.
\end{eqnarray}
In this way, the particle anti-particle $q$-excess takes the form
\begin{eqnarray}
	n_{r^{+}}^{q} - n_{r^{-}}^{q} &=& \frac{gB_{q}}{\lambda^{3/2}}\left( e_{q}^{-\beta m}\right)^{\frac{5-3q}{2}} \left[\left(  e_{q}^{\alpha_q\mu_{r^{+}}}\right)^{\frac{5-3q}{2}}\right.
	\nonumber \\
	&-& \left. \left(e_{q}^{-\alpha_q\mu_{r^{+}}}\right)^{\frac{5-3q}{2}}\right]~,
\end{eqnarray}
where $\alpha_q =\beta/[1+(q-1)\beta m]$. The limit $q\rightarrow 1$ leads to $B_{q}\rightarrow 1$, $\alpha_q\rightarrow \beta$ and $e_{q}^{x}\rightarrow e^{x}$, therefore, the standard approach is recovered, as below:
\begin{eqnarray}
	n_{r^{+}}^{1} - n_{r^{-}}^{1} = 2g\left(\frac{mT}{2\pi}\right)^{3/2}e^{-\beta m}\sinh(\beta \mu_{r^{+}})~. 
\end{eqnarray}

Since we have not yet determined a version of Eq. (\ref{q-density}) for relativistic fermions ($m_{i}\ll T$), this is our task from now on. As far as we knowledge, the theoretical outcome for the non-Gaussian particle anti-particle excess, as we will show here, has not been reported earlier. According to the result seen in \ref{appendixB}, the particle anti-particle $q$-excess reads as
\begin{eqnarray} \label{q-excess}
	n_{r^{+}}^{q} - n_{r^{-}}^{q} &=& \frac{g}{2\pi^2}T^{3}\left\lbrace \frac{2}{\phi_{q}^{3}(a)}\left[ I_{q}^{*}(e_{q}^{a}) - I_{q}^{**}((e_{q}^{a})^{2})\right] \right.
	\nonumber \\
	&-& \left. 
	\frac{2}{\phi_{q}^{3}(-a)}\left[ I_{q}^{*}(e_{q}^{-a}) - I_{q}^{**}((e_{q}^{-a})^{2})\right] \right\rbrace \;,
\end{eqnarray}
where  
\begin{eqnarray} 
	I_{q}^{*}(z) &=& \frac{z}{2(q-1)^{2}}\left[ \Phi(z, 1, 4-3q)-2\Phi(z, 1, 3-2q) \right.
	\nonumber \\
	&+& \left.
	\Phi(z, 1, 2-q) \right]\;,
\end{eqnarray}
\begin{eqnarray} 
	I_{q}^{**}(z) &=& \frac{z}{2(q-1)^{2}}\left[ \Phi(z, 1, (5-3q)/2)-2\Phi(z, 1, 2-q) \right.
	\nonumber \\
	&+& \left.
	\Phi(z, 1, (3-q)/2) \right]\;,
\end{eqnarray}
and
\begin{eqnarray}
	\phi_{q}(\pm a) =\frac{1}{1\pm (1-q)a}\;.
\end{eqnarray}
In the expressions above, we defined $a=\beta\mu_{r^{+}}$. Note that $\Phi(z, s, b)$ is the Hurwitz-Lerch zeta function or Lerch transcendent function. It is worth noting that the results for $I_{q}^{*}(z)$ and $I_{q}^{**}(z)$ are real only if $|z|<1$ and that this function diverges when $\Phi(z, 1, 0)$. Similarly to Eq. (\ref{q-number-pho}), the $q$-parameter in $I_{q}^{*}(z)$ obeys the same constraints being the most suitable range is also $0<q<4/3$. Besides, the interval $4/3<q<3/2$ yields $I_{q}^{*}(z)<0$, and this is not allowed. On the other hand, the $q$-parameter in $I_{q}^{**}(z)$ is constrained according to $q\neq 5/3, q\neq 3 \;{\rm and}\; q\neq 2$. In such a way, we have three possible scenarios: $0<q<5/3$, $5/3<q<2$, and $2<q<3$. Here, the interval $5/3<q<2$ leads to $I_{q}^{**}(z)<0$, which is not a result of physical interest. For reasons already discussed previously, the values $q<0$ are not appropriate. Therefore, the interval most convenient is $0<q<5/3$.

The limit $q\rightarrow 1$ leads to $e_{q}^{z}\rightarrow e^{z}$, $I_{q}^{*}(z)\rightarrow Li_{3}(z)$, $I_{q}^{**}(z)\rightarrow Li_{3}(z)/4$ and $\phi_{q}=1$, where $Li_{s}(z)$ is the polylogarithm function. Hence, we obtain
\begin{eqnarray}
	n_{r^{+}}^{1} - n_{r^{-}}^{1} &=& \frac{g}{2\pi^2}T^3\left[ 2Li_{3}(e^a)-\frac{1}{2}Li_{3}(e^{2a})\right.
	\nonumber\\
	&-&\left.2Li_{3}(e^{-a})+\frac{1}{2}Li_{3}(e^{-2a}) \right]\;.
\end{eqnarray}
Considering now the limit for a non-degenerate gas, i.e. $a\approx 0$, the result above takes the form as is well known in the literature \cite{Padmanabhan,Houjun}
\begin{eqnarray}
	n_{r^{+}}^{1} - n_{r^{-}}^{1} = \frac{g}{6\pi^2}T^3\left[\pi^{2}\frac{\mu_{r^{+}}}{T}+\left(\frac{\mu_{r^{+}}}{T}\right)^3\right]\;,
\end{eqnarray}
where we performed an expansion up to the third order of $a$.  

\section{Conclusion} \label{conclusion}

In the literature, few studies on the generalization of the Saha's ionization using non-Gaussian statistics have been carried out in recent years. In particular, no non-Gaussian effect on binding energy has been reported so far. In this present paper, we showed that, unlike the standard case, there is a non-Gaussian chemical equilibrium condition for each energy regime, i.e., one for the non-relativistic limit and the other for the relativistic one. Next, the non-Gaussian Saha's ionization has been calculated considering the chemical equilibrium condition for the non-relativistic limit so that the final expression was written independently of the chemical potential of the species, which makes it stand out from other approaches found in the literature. This result suggests a generalized form for the binding energy of the hydrogen atom. This new binding energy depends on the $q$-parameter and the temperature. We analyzed the behavior of the binding $q$-energy and showed that for high temperatures it approaches its conventional value, and at low temperatures, it becomes too large for $a>0$ and too small for $a<0$. In summary, we demonstrated that to yields tiny shifts on the binding energy, the $a$-parameter need be quite small, i.e., $a\sim 10^{-16}$. Since the binding $q$-energy is a ground state shift, negative values of the $a$-parameter are mandatory for accessing other hydrogen energy levels. We showed that for $n\geq 2$, the $a$-parameter decreases linearly as the temperature increases. We also discussed the behavior of the $a$-parameter for some hydrogen energy levels. The results indicate that at low temperatures the behavior of the $a$-parameter shifts subtly mainly for $n>3$ and displays tiny differences as the temperature rises. Besides, we find out a non-Gaussian version of the baryons density via the baryon-to-photon ratio. The result obtained was expressed in terms of the polygamma function in the validity range for baryon $q$-density $0<q<4/3$. As applications, we investigated the non-Gaussian effects of the Saha's ionization on the deuterium bottleneck, primordial hydrogen recombination, and the particle anti-particle excess. Concerning the deuterium bottleneck, we obtained a non-Gaussian version of the deuterium-to-baryon ratio and showed that when $q\approx 1.152$ the temperature at which the BBN starts is $T_{\rm BBN}\approx 0.062~{\rm MeV}$, the same value given by standard approach. This result suggests that this value of the factor $q$ is in the equilibrium range in the non-Gaussian context. We also analyzed other temperature scenarios and noted that $T_{\rm BBN}\propto 1/q$. Therefore, from our estimates, the $q$-parameter plays a role in controlling the temperature of the thermal bath. Regarding recombination, we estimated the value $q\approx 1.209$ so that the equilibrium fractional ionization temperature is about $T_{\rm dec}\approx 0.321~{\rm eV}$, and also noticed that $T_{\rm dec}\propto 1/q$. We used the binding $q$-energy (assuming $n=2$) to estimate the $a$-parameter at the beginning and end of recombination, and when $x_e\approx 0.5$. We also estimated a upper bound for the $a$-parameter assuming $n\rightarrow \infty$ and $T\approx\varepsilon_{0}$. Moreover, we obtained generalized expressions for the particle anti-particle excess. For the non-relativistic limit, we used the number $q$-density and for the relativistic one, we presented a result in terms of the Hurwitz-Lerch zeta function whose interval most suitable is $0<q<5/3$. Finally, we showed that both return to their usual forms in the limit $q\rightarrow 1$.

\begin{acknowledgements}
	The authors are grateful to the Brazilian agency CAPES for financial support. FCC was supported by CNPq/FAPERN/PRONEM.
\end{acknowledgements}

\appendix

\section{\label{appendixA} Exact solution for photon density}

The photons $q$-density is given by [see Eq. (\ref{intq})]
\begin{eqnarray} 
n_{\gamma}^{q} = \frac{8\pi}{(2\pi\beta)^{3}}C_q~,
\end{eqnarray}
where
\begin{eqnarray} 
C_q=\int_{0}^{\infty}\frac{x^{2}dx}{e_{2-q}^{x}-1}~.
\end{eqnarray}
Using the property $e_{q}^{-x}e_{2-q}^{x}=1$ and the definition of a geometric series is easy to show that
\begin{eqnarray} 
C_q= \sum_{n=1}^{\infty}\int_{0}^{\infty}x^{2}\left[1-(1-q)x\right]^{\frac{n}{1-q}}dx \;.
\end{eqnarray}
The integral above can be solved using integration by parts. Thus, we obtain the following result:
\begin{eqnarray} 
C_q = \sum_{n=1}^{\infty} \frac{2}{(n-q+1)(n-2q+2)(n-3q+3)}\;.
\end{eqnarray}
We can write this sum in terms of known functions, namely
\begin{eqnarray}
C_q = \frac{2\psi^{(0)}(3-2q)-\psi^{(0)}(4-3q)-\psi^{(0)}(2-q)}{(q-1)^2}~,
\nonumber\\
\end{eqnarray}
where $\psi^{(n)}(z)$ is the polygamma function.

Let us now analyze the limit when $q\rightarrow 1$. In this case, we have that
\begin{eqnarray} 
\lim_{q \to 1}C_q = -\psi^{(2)}(1)=2\zeta(3)\;.
\end{eqnarray}
Then, the standard result is recovered, as follow:
\begin{eqnarray} 
n_{\gamma}^{1} = \frac{2\zeta(3)}{\pi^2}T^{3}\;,
\end{eqnarray}
where $\zeta(z)$ is the well known Riemann's zeta function.

\section{\label{appendixB} Exact solution for particle anti-particle excess}

The particle anti-particle excess for relativistic fermions is calculated as follow: 
\begin{eqnarray} 
n_{r^{+}}^{q} - n_{r^{-}}^{q} = \frac{4\pi g}{(2\pi)^{3}}T^{3}M\;,
\nonumber\\
\end{eqnarray}
where, for simplicity, we defined $M=M'-M''$ so that
\begin{eqnarray} 
M' = \int_{0}^{\infty}\frac{x^{2}dx}{{e_{2-q}^{x-a}}+1} 
\end{eqnarray}
and
\begin{eqnarray}
M'' = \int_{0}^{\infty}\frac{x^{2}dx}{{e_{2-q}^{x+a}}+1}\;,
\end{eqnarray}
with $x=\beta E$. Let us first solve the integral $M'$. In order to get the result, it considers the following identity:
\begin{eqnarray}
\frac{1}{{e_{2-q}^{x}}+1} = \frac{1}{{e_{2-q}^{x}}-1}-\frac{2}{({e_{2-q}^{x}})^{2}-1}\;.
\end{eqnarray}
That way, $M'$ is redefined to $M'=M^{*}-2M^{**}$, where 
\begin{eqnarray} 
M^{*} = \int_{0}^{\infty}\frac{x^{2}dx}{{e_{2-q}^{x-a}}-1}
\end{eqnarray}
and
\begin{eqnarray} 
M^{**} = \int_{0}^{\infty}\frac{x^{2}dx}{({e_{2-q}^{x-a}})^{2}-1}\;. 
\end{eqnarray}
Having assigned these definitions, we are in a position for solving the integral $M'$. Let us firstly solve $M^{*}$ and then $M^{**}$. Using the property $e_{q}^{-x}e_{2-q}^{x}=1$ again and the definition of a geometric series is straightforward to show that
\begin{eqnarray} 
M^{*} = \frac{1}{\phi_{q}^{3}(a)}\sum_{n=1}^{\infty}\left(e_{q}^{a}\right)^{n} \int_{0}^{\infty}x^{2}\left[1-(1-q)x\right]^{\frac{n}{1-q}}dx\;,
\nonumber\\
\end{eqnarray}
where we used the $q$-algebra properties and defined
\begin{eqnarray} 
\phi_{q}(a) = \frac{1}{1+(1-q)a}\;.
\end{eqnarray}
The integral above has already been resolved in the previous Appendix. So, we obtain
\begin{eqnarray} 
M^{*} = \frac{2}{\phi_{q}^{3}(a)}\sum_{n=1}^{\infty}\frac{\left(e_{q}^{a}\right)^{n}}{(n-q+1)(n-2q+2)(n-3q+3)}\;.
\nonumber\\
\end{eqnarray}
This sum can also be written in terms of known special functions as long as $|e_{q}^{a}|<1$. Thus, we find
\begin{eqnarray} 
M^{*} = \frac{2}{\phi_{q}^{3}(a)}I_{q}^{*}(e_{q}^{a})\;,
\end{eqnarray}
where
\begin{eqnarray}
I_{q}^{*}(e_{q}^{a}) &=& \frac{e_{q}^{a}}{2(q-1)^{2}}\left[ \Phi(e_{q}^{a}, 1, 4-3q)-2\Phi(e_{q}^{a}, 1, 3-2q)\right.
\nonumber\\
&+&\left.\Phi(e_{q}^{a}, 1, 2-q) \right]\;, 
\end{eqnarray}
with $\Phi(z, s, b)$ being the Lerch transcendent function.

Likewise, we can solve the integral $M^{**}$. Hence, we have
\begin{eqnarray} 
M^{**} = \frac{1}{\phi_{q}^{3}(a)}\sum_{n=1}^{\infty}\frac{\left(e_{q}^{a}\right)^{2n}}{(2n-q+1)(n-q+1)(2n-3q+3)}\;,
\nonumber\\
\end{eqnarray}
or more compactly as
\begin{eqnarray} 
M^{**} = \frac{1}{\phi_{q}^{3}(a)}I_{q}^{**}((e_{q}^{a})^{2})\;,
\end{eqnarray}
with
\begin{eqnarray}
I_{q}^{**}((e_{q}^{a})^{2}) &=& \frac{(e_{q}^{a})^{2}}{2(q-1)^{2}}\left[ \Phi((e_{q}^{a})^{2}, 1, (5-3q)/2)\right.
\nonumber\\
&-&2\left.\Phi((e_{q}^{a})^{2}, 1, 2-q)+\Phi((e_{q}^{a})^{2}, 1, (3-q)/2) \right]\;, 
\nonumber\\
\end{eqnarray}
since $|(e_{q}^{a})^{2}|<1$. With these results, we can conclude that
\begin{eqnarray}
M' =  \frac{2}{\phi_{q}^{3}(a)}\left[ I_{q}^{*}(e_{q}^{a}) - I_{q}^{**}((e_{q}^{a})^{2})\right] \;.
\end{eqnarray}

Correspondingly $M''$ is obtained replacing $a\rightarrow -a$. Therefore,
\begin{eqnarray}
M'' =  \frac{2}{\phi_{q}^{3}(-a)}\left[ I_{q}^{*}(e_{q}^{-a}) - I_{q}^{**}((e_{q}^{-a})^{2})\right] \;.
\end{eqnarray}
And finally
\begin{eqnarray}
M &=& \frac{2}{\phi_{q}^{3}(a)}\left[ I_{q}^{*}(e_{q}^{a}) - I_{q}^{**}((e_{q}^{a})^{2})\right] 
\nonumber\\
&-& \frac{2}{\phi_{q}^{3}(-a)}\left[ I_{q}^{*}(e_{q}^{-a}) - I_{q}^{**}((e_{q}^{-a})^{2})\right] \;. 
\end{eqnarray}
Thus, the particle anti-particle excess for relativistic fermions in the Tsallis framework can be written as in Eq. (\ref{q-excess}).

\end{document}